\documentclass[11pt]{iopart}
\usepackage{iopams}  
\usepackage{textcomp} 
\usepackage{graphicx}
\usepackage[numbers,square,sort&compress]{natbib} 
\usepackage{color}

\def\stef#1{#1}

\begin{document}

\title[]{Femtosecond envelope of the high-harmonic emission from ablation plasmas}

\author{S Haessler$^{1,2}$, L. B. Elouga Bom$^3$, O. Gobert$^1$, J.-F. Hergott$^1$, F. Lepetit$^1$, M. Perdrix$^1$, B Carr\'e$^1$, T. Ozaki$^3$ and P Sali\`eres$^1$}

\address{1 CEA Saclay, IRAMIS, Service des Photons, Atomes et Mol\'ecules, 91191 Gif-sur-Yvette, France.}
\address{2 Photonics Institute, Vienna  University of Technology, Gu\ss hausstra\ss e 27/387, A-1040 Vienna, Austria.}
\address{3 Institut National de la Recherche Scientifique, Centre Energie, Mat\'eriaux et T\'el\'ecommunications, 1650 Lionel-Boulet, Varennes, Qu\'ebec J3X 1S2, Canada.}
\ead{pascal.salieres@cea.fr}

\begin{abstract}
We characterize the temporal profile of the high-order harmonic emission from ablation plasma plumes using cross-correlations with the infrared (IR) laser beam provided by 2-photon harmonic+IR ionization of rare gas atoms. We study both non-resonant plasmas (lead, gold and chrome) and resonant plasmas (indium and tin), i.e. plasmas presenting in the singly-charged ions a strong radiative transition coinciding with an harmonic order. The cross-correlation traces are found very similar for all harmonic orders and all plasma targets. The recovered harmonic pulse \stef{durations are very similar to the driving laser, with a tendency towards being shorter,} demonstrating that the emission is a \stef{directly} laser-driven process even in the case of resonant harmonics. This provides valuable input for theories describing resonant harmonic emission and opens the perspective of a very-high-flux tabletop \textsc{xuv} source for applications.

\end{abstract}

\submitto{\JPB}

\section{Introduction}

High-order harmonic generation (HHG) occurs when an intense short laser pulse is focused into a nonlinear medium. It was first studied in rare gas atoms and was shown to produce a radiation in the extreme Ultraviolet (XUV) with ultrashort duration. First, the femtosecond envelope of the emission was characterized \cite{Glover1996,Bouhal1997,Mauritsson2004xfrog,Mairesse2005HHSPIDER}. Then the attosecond substructure was studied, demonstrating the possibility to generate isolated \cite{Hentschel2001Attosecond,Sansone2006,Feng2009generation} or trains of \cite{Paul2001Observation,Mairesse2003Attosecond,Kim2010aptfrog} attosecond pulses. The femtosecond/attosecond nature of the harmonic emission, along with its short wavelength, excellent coherence and high brilliance, provide a strong motivation for applications, some of which have already been demonstrated in atomic and molecular spectroscopy and solid-state physics \cite{Krausz2009}. Despite its potential, many other applications remain unexplored because of the difficulty to handle and manipulate the harmonics, due partially to the quality of \textsc{xuv} optics, as well as to the low harmonic intensity. 
 
Instead of rare gas atoms, different generating media have been tested in order to increase the conversion efficiency. Recently, attosecond pulses have been successfully generated in molecules. Aligned linear molecules allow a coherent control of the attosecond emission \cite{Boutu2008Coherent}, or a tomographic reconstruction of the radiating orbital \cite{Itatani2004Tomographic,Haessler2010tomo}, but unfortunately no real improvement of the efficiency. Due to their high ionization potentials, singly-charged ions are interesting candidates since they can stand higher laser intensity and thus can potentially generate both more intense radiation and higher harmonic orders. Such ions can be produced through ionization of rare gases: \textsc{hhg} in Ar$^+$ was shown to extend the cutoff to 250 eV \cite{Gibson2004}. Another possibility is to produce a lowly-ionized ablation plasma by focusing a long ($\sim100\:$ps) \textsc{ir} laser pulse at $I\sim10^{10}\:$W/cm$^2$ onto the surface of a solid target material. After letting the ablation plasma expand for $\sim100\:$ns, a strong femtosecond laser pulse, propagating parallel to the target surface at a distance of $\sim100\:\mu$m, may generate high-order harmonics of the driving laser in the plasma. The plasma \textsc{hhg} medium predominantly contains singly charged ions and small amounts of neutral atoms and higher charged ions~\cite{Elouga2008Correlation}. The plasma is under-dense, i.e. the associated plasma-frequency of the collective free electron oscillation is far below the laser frequency $\omega_0$ such that the plasma is completely transparent.

The idea to use such a medium came up quite early in the quest for an optimized light source based on \textsc{hhg} \cite{Akiyama1992,Wahlstrom1995}. At the time, the aim was to use the larger ionization potential of alkali ions as compared to rare-gas atoms to to extend the \textsc{hhg} plateau to very high photon energies. However, no real improvement was obtained, presumably due to the difficult plasma conditions. In 2005, the method re-attracted attention when Ganeev et al. \cite{Ganeev2005boron,Ganeev2005Silver} generated harmonics up to the 63rd order using boron and silver targets, and reported conversion efficiencies in the plateau region as high as $\sim10^{-5}$. Such values had been obtained with gas harmonics after serious efforts optimizing phase matching \cite{Hergott2002Extreme}, whereas they seemed to be `easily' achievable with ablation plasmas. Soon after, this value was surpassed for single harmonics by exploiting a resonant enhancement effect, leading to a $\sim10^{-4}$ conversion efficiency for harmonic 13 and 17 in indium and tin plasmas, respectively \cite{Ganeev2006Strong,Suzuki2006Anomalous}. Since recently, nano-particle~\cite{Ganeev2009Comparison} and fullerene~\cite{Ganeev2009} containing targets are studied, because they can be highly polarizable and offer the possibility to tune plasmon resonances to a frequency of choice by varying the nano-particle size. A review of this approach as well as the general state-of-the-art of \textsc{hhg} in ablation plasma plumes is given in \cite{Ganeev2007}.

Up to very recently, experiments focused on the maximization of generated \textsc{xuv} photon numbers. Few systematic studies have been conducted to shed light on the physical process responsible for the \textsc{xuv} emission and the reported high efficiency. The best available support for the presumption that the generation mechanism resemble the three-step process of the gas harmonics had been the observation of a similarly rapid drop of photon yield with increasing ellipticity of the driving laser that was observed for non-resonant~\cite{Ganeev2005boron,Ganeev2005Silver} as well as resonant~\cite{Suzuki2006Anomalous} \textsc{hhg} in ablation plasma. \stef{This is a strong indication that it is electron recollision that is responsible for the \textsc{xuv} emission. Very recently, signatures of the short and long quantum paths known from \textsc{hhg} in neutral gases have been observed in the spectrally resolved  far-field spatial profile \cite{Ganeev2011quantumpath} of harmonics from a \textit{non-resonant} aluminum plasma. However, a property particularly important both for its fundamental and applied aspects had not yet been studied: the ultrashort \textsc{xuv} pulse duration. Although the analogy to \textsc{hhg} in neutral gases would indicate a femtosecond/attosecond character at least in the non-resonant cases, one should not forget that the \textsc{hhg} medium conditions in the case of plasmas are much more complicated: one has a mixture of different ion charge states and quite possibly clusters, and it is not obvious which species contribute significantly to the signal. Moreover, the dispersion caused by the much higher free electron density may sensitively influence the phase matching and the XUV and laser propagation. These issues concern the macroscopic aspect of \textsc{hhg} and will impact on the \textsc{xuv} temporal structure. The case of resonant plasmas raises even more questions: both the mechanism for the microscopic response and the role of the strong medium dispersion on the macroscopic emission are not clear. Temporal characterization is thus in dire demand, and only when the femtosecond/attosecond character has been shown experimentally, \textsc{hhg} in ablation plasma can indeed be considered a promising source of intense ultrashort coherent \textsc{xuv} pulses for applications.}

We thus set up an experiment in Saclay to measure these temporal properties, which was done in two experimental campaigns. The characterization of the attosecond emission from a chromium \textit{non-resonant} plasma was reported in \cite{Elouga2011chrome}. We measured a very low group delay dispersion allowing the generation of near-transform-limited 300~as pulses from the spectral region covering harmonics 11 to 19 . The investigation of the temporal properties of the \textsc{hhg} from a tin \textit{resonant} plasma is presented in \cite{Haessler2011submitted}. We found very similar femtosecond envelopes for the resonant and non-resonant harmonic orders demonstrating that the enhancement in tin plasmas is a field-driven process. However, the phase locking between the resonant and the neighboring harmonics was found very degraded, indicating strong phase distortions in the emission process and a blurred attosecond sub-structure. In this paper, we give a full account of our measurements of the femtosecond envelope of the emission from a variety of plasma targets. In Section \ref{sec:setup}, we describe our experimental setup and discuss the experimental conditions in Section \ref{sec:exp}. In Section \ref{sec:plasmanonreso}, we present our results obtained in non-resonant ablation plasmas (lead, gold and chrome) and in Section \ref{sec:plasmareso}, those in resonant plasmas (indium and tin). In Section \ref{sec:discussion}, we discuss the duration measured for the cross-correlations, the estimated duration for the \textsc{xuv} emission and finally conclude.

\section{Experimental setup}
\label{sec:setup}
\begin{figure}[tb]
\centering\includegraphics[width=0.55\textwidth]{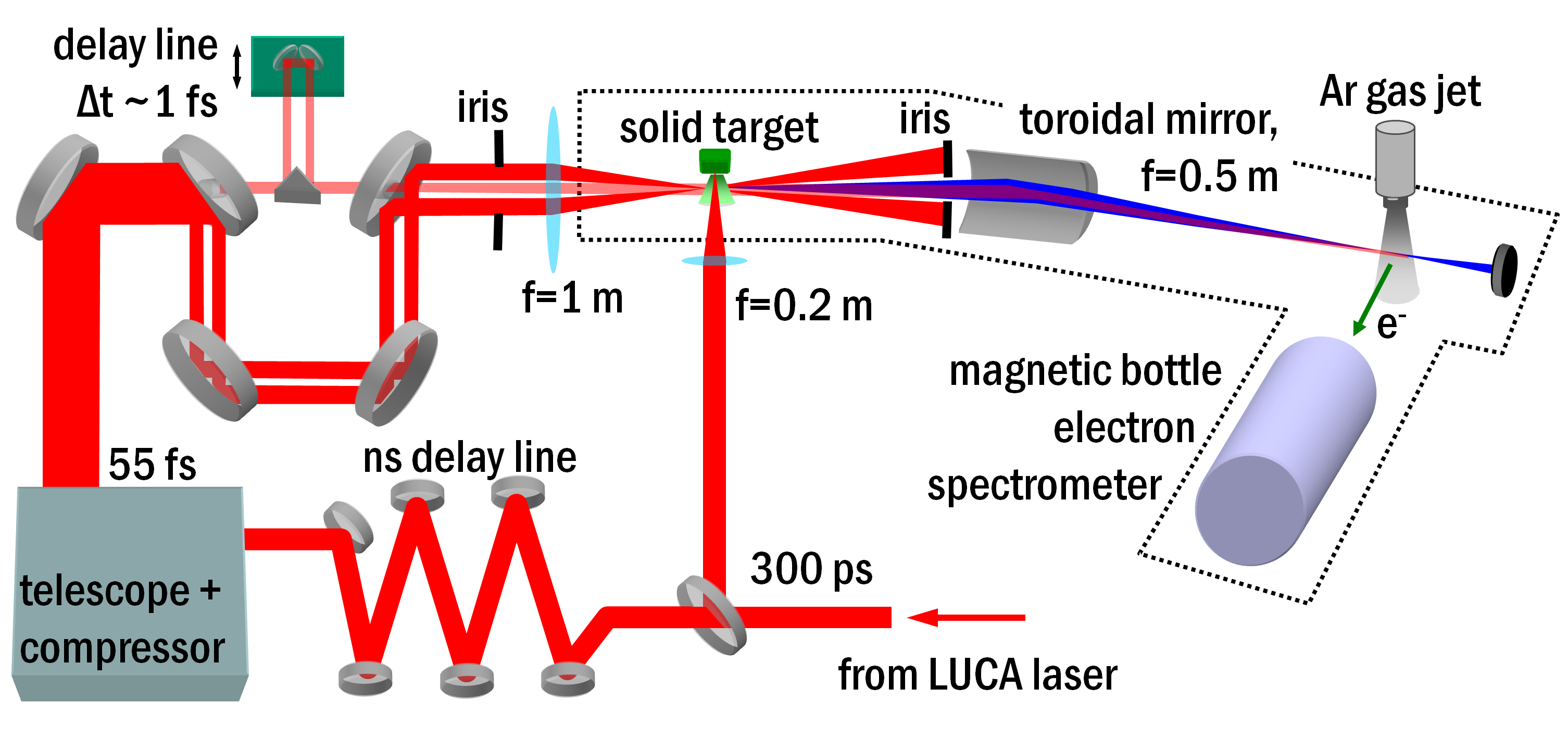}
\caption{Experimental set-up for the generation and characterization of the femtosecond emission from an ablation plasma.}
\label{Fig:setup}
\end{figure}
The experimental setup is schematically shown in Fig. \ref{Fig:setup}. First, we separated the uncompressed Ti:sapphire laser beam (\stef{20~Hz,} 300~ps) from the ``LUCA'' laser system into two beams, using a beam splitter. One of these beams (\stef{heating} pulse) is loosely focused \stef{(beam diameter $\lesssim10\:$mm, pulse energy $\lesssim10\:$mJ)} at normal incidence onto a solid target \stef{($\approx20\:$mm before the beam focus, beam spot diameter $\lesssim1\:$mm, intensity \mbox{$\sim10^9$--$10^{10}\:$W/cm$^2$)}} to create the lowly-ionized plasma. The other uncompressed beam is first sent through an 80-ns delay line, and then through a compressor. At the exit of the compressor, the 55-fs laser pulse is further separated into two beams using a mirror with \stef{an 8-mm} through hole at the center. The external annular part of the beam (generating beam\stef{, outer diameter $\approx17\:$mm, pulse energy $\approx6\:$mJ }) is focused parallel to the target surface into the plasma to generate the high-order harmonics. The round central part of the beam (dressing beam\stef{, diameter $\lesssim4\:$mm}), which is at least two orders of magnitude less intense than the external part, passes through the hole of the mirror and enters a very precise delay line equipped with a piezoelectric translation stage. It is then recombined with the external part of the beam through a second holed mirror. After harmonic generation, a \stef{4-mm-}iris blocks the generating beam while the more collimated harmonic beam together with the dressing beam are refocused by a toroidal mirror into a detection argon jet placed in the source volume of a magnetic bottle electron spectrometer (MBES). The time of flight, and thus the energy, of the electrons produced by photo-ionization of the target gas is measured with an oscilloscope.

Sidebands, created in photoelectron spectra by two-photon \textsc{xuv}$\pm$\textsc{ir} absorption provide a means to record cross-correlation traces of an \textsc{ir} probe pulse and the \textsc{xuv} pulses~\cite{Glover1996,Bouhal1997,Bouhal1998,Norin2002,Mauritsson2005}. Indeed the sideband intensity can be written as:
\begin{equation}
	 I_\mathrm{SB}(\tau)=\int_{-\infty}^\infty I_\mathrm{\textsc{xuv}}(t) \, f[I_\mathrm{IR}(t-\tau)] \, \mathrm{d}t.
	 \label{eq:plasma:XcoI}
\end{equation} 
The function $f[I_\mathrm{IR}]$ is well approximated by a power law $\propto {I_\mathrm{IR}}^\alpha$ at moderate \textsc{ir} intensity \cite{Bouhal1997}. For very low intensities $\lesssim1\times10^{11}\:$W/cm$^2$, $\alpha$ converges to 1.
From the temporal width of a 1st order sideband, one can then estimate the average temporal width of the two adjacent odd harmonic orders if the \textsc{ir} probe pulse is known. For a proper de-correlation, the data has to be of exceptional quality, which is usually not the case. One thus typically assumes Gaussian temporal profiles for both the \textsc{xuv} and \textsc{ir} pulse, and estimates the FWHM duration, $\tau_\mathrm{XUV}$, of the \textsc{xuv} as
\begin{equation}
	 \tau_\mathrm{XUV}  = \sqrt{ \tau_\mathrm{SB}^2-\tau_\mathrm{IR}^2/\alpha},
	 \label{eq:plasma:Xcotau}
\end{equation}
where $\tau_\mathrm{SB}$ and $\tau_\mathrm{IR}$ are the measured sideband temporal width and the known \textsc{ir} probe pulse duration, respectively. In addition to this estimate, the temporal shape of the cross-correlation traces contains information about the pulse shapes (i.e. deviations from the assumed Gaussian profile) of \textsc{ir} and \textsc{xuv}. 

\begin{figure}[tb]
\centering\includegraphics[width=0.45\textwidth]{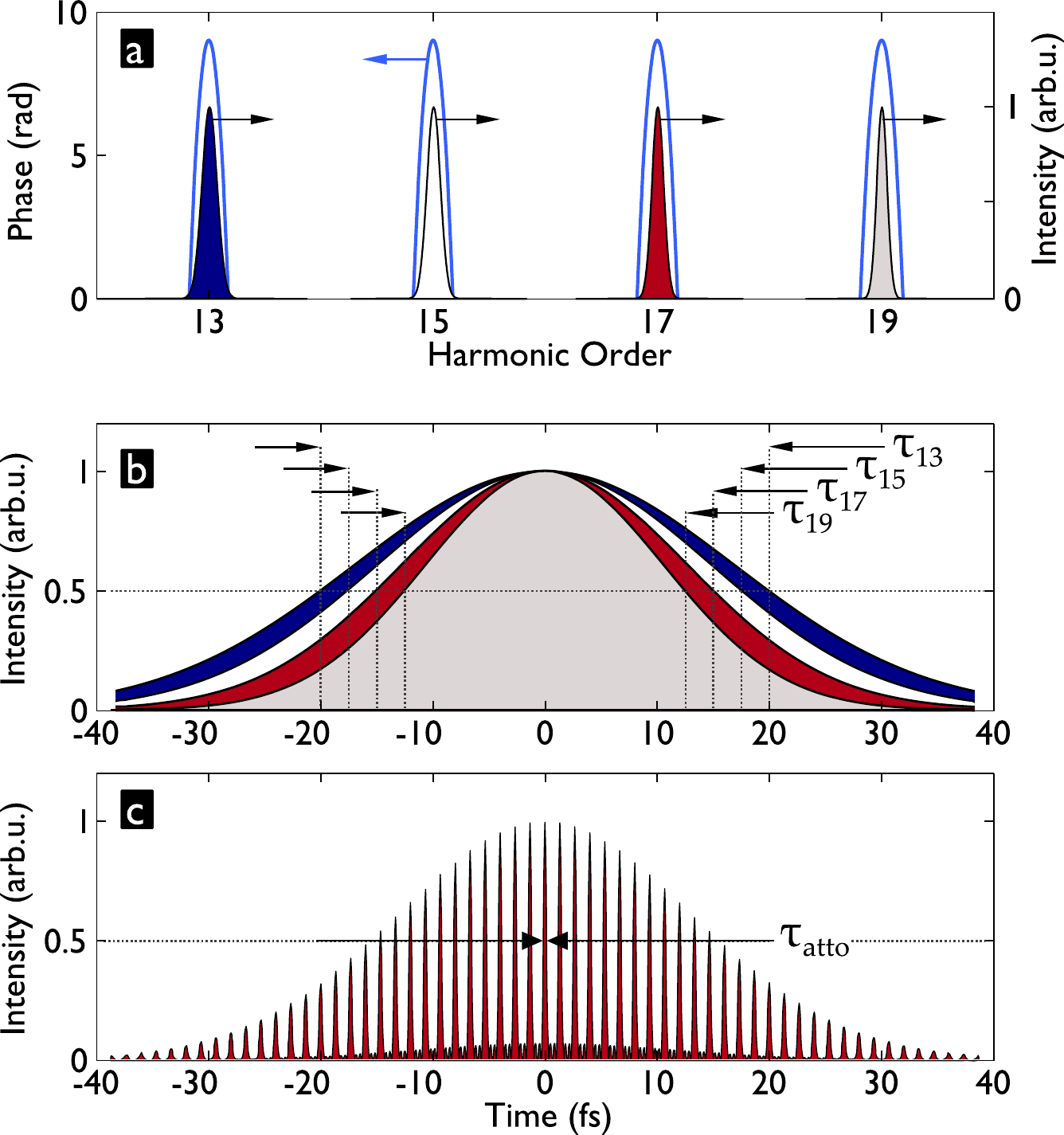}
\caption{\stef{Characterizing the harmonic emission on different time scales: as an example, the spectral intensity and phase of four harmonics of an 800-nm laser is shown in panel (a). Fourier transforming each harmonic peak individually, the pulse durations corresponding to each harmonic order are found as shown in panel (b): the 13th, 15th, 17th and 19th harmonic have Gaussian temporal intensity envelopes with FWHM durations of $\tau_{13, 15, 17, 19}=40,\:35,\:30$ and 25~fs, respectively. Coherently superposing all these pulses, or equivalently Fourier transforming the full spectrum comprising four harmonic orders, leads to the attosecond pulse train shown in panel (c). For RABBIT, the harmonics are considered as having sharp frequencies and phases (i.e. zero spectral width). This yields an attosecond pulse duration, $\tau_\mathrm{atto}=300\:$as, representative for the attosecond pulses around the maximum of the pulse train envelope.}}
\label{fig:APT}
\end{figure}
The setup is thus the same as for RABBIT (Reconstruction of Attosecond Beating By Interference of two-photon Transitions) measurements \cite{Elouga2011chrome,Paul2001Observation,Mairesse2003Attosecond}. \stef{The principle difference between RABBIT measurements and those described in the present manuscript is simply that RABBIT is an interferometric cross-correlation, where the XUV-IR delay is scanned with sub-laser-cycle steps over only a few laser periods (say, 10 fs), whereas here, we report intensity cross-correlation measurements,  the scanned delay range will cover the complete temporal overlap of the \textsc{xuv} and the \textsc{ir} probe pulse envelopes ($\sim200\:$fs). Since only the temporal envelope of the sideband-signal is analyzed,} interferometric stability is not necessary. In our case, where the \textsc{ir} \textsc{hhg} driving beam and \textsc{ir} probe beam interfere in the \textsc{hhg} gas jet, it is in fact undesirable, since it would lead to an $\omega_0$-modulation of the cross-correlation signal that would have to be removed afterwards. Our measurements were thus actually done with efforts to destabilize the RABBIT-interferometer, which is obviously rather easy to do. Programming the piezo-electric transducer to do some jitter during the data acquisition would have been an option but having the air-conditioning blow onto the uncovered optical table already provided the required perturbation.

\stef{Figure \ref{fig:APT} illustrates the different time-scales on which the harmonic emission is characterized by RABBIT or cross-correlation measurements. In the present manuscript, we discuss the pulse-duration of \emph{individual harmonic orders} (see Fig. \ref{fig:APT}b), which could be brought on target with a relatively narrow band filter (few 100 meV) that selects only a single harmonic. RABBIT, on the other hand, addresses the average duration of the attosecond pulses in the pulse-train (which has a femtosecond envelope), formed by the \emph{coherent superposition of several harmonic orders} (see Fig. \ref{fig:APT}c). This could be produced on target when using some band-pass filter (10s of eV width) that selects several harmonics. More details can be found in \cite{Varju2005}.}

\section{Experimental conditions}
\label{sec:exp}

A delicate element in this setup is the $\approx80\:$ns delay line. The passage through the compressor imposes already a delay of $\approx20\:$ns, leaving $60\:$ns, corresponding to 18 m of propagation to be added. The precise length is not very important---as long as the delay between plasma creation and \textsc{hhg} is $\gtrsim50\:$ns, the ablation plasma conditions should be stable~\cite{Ganeev2006Harmonic}. In an attempt to avoid pulse distortions due to very long propagation in air, we installed an optical system which imaged the entrance point of the delay line onto the exit point, localized after four passages over 4.5 m, most \stef{of} which was maintained in vacuum. Unfortunately, hot-spots in the laser beam profile, occurring somewhere before the laser amplifier chain exit, were imaged at points within the 18 m of the delay line. Some of these fell onto the folding mirrors of the delay line, regularly leaving burn marks, which then introduced distortions to the beam profile, creating even more hot spots. The experiments were thus complicated by a distorted spatial beam quality. The pulse duration was barely affected, though, and was measured to be $\approx55\:$fs at the entrance of our setup with an auto-correlator and at the exit (after the MBES) with SPIDER.

We were limited to a generating intensity of  $\lesssim2\times10^{14}\:$W/cm$^2$ at the laser beam focus in the \textsc{hhg} chamber, estimated via the attochirp of high harmonics generated in argon gas by the laser beam that passed the delay line. \stef{The attochirp, which is inversely proportional to the generating intensity \cite{Mairesse2003Attosecond}, was measured with RABBIT for several diameters of the generating beam after replacing the metal target by an argon gas jet. A comparison with attochirp values from SFA simulations \cite{Lewenstein1995, Mairesse2003Attosecond} then allows determining the effective generating intensity up to $\pm10:$\% . The maximum intensity} was reached when the beam diameter, which is $\approx35\:$mm at the compressor exit, was cut down to $\approx17\:$mm just before the focusing lens. With this focusing geometry and an optical setup without the ns-delay line, we usually need two to three times less pulse energy to reach $2\times10^{14}\:$W/cm$^2$ at focus. Increasing the beam diameter and thus the pulse energy led to a decreasing intensity at focus, which clearly hints at wave front distortions suffered by the laser beam in the ns-delay line.

In earlier experiments (see e.g. \cite{Ganeev2005Silver,Elouga2007Influence}, where a silver target was used) for which the intriguingly high conversion efficiencies had been reported, the generating intensities were at least factor 2 above the barrier suppression intensity of the singly charged ions \cite{Ilkov1992}:
\begin{equation}
I_\mathrm{BS}[\mathrm{W/cm}^{2}]= 4\times10^9  \,\frac{I_\mathrm{p} [\mathrm{eV}]^4}{Z^2},
\label{eq:IBS}
\end{equation}
where $I_\mathrm{p}$ is the ionization potential and $Z$ is the effective charge of the screened nucleus, i.e. $Z=1$ for the neutrals, $Z=2$ for singly charged ions, and so forth. Also, the \stef{maximized} cutoff harmonic order $H_\mathrm{c}$, obtained for a range of targets and an \mbox{800-nm} driving laser, could be fitted with the relation $H_\mathrm{c} \approx 4 I_\mathrm{p} -32.1$ \cite{Ganeev2007}. Using the classical cutoff law for HHG \cite{Schafer1993Above,Corkum1993Plasma}, one finds that these cutoff positions require 2--3 times the barrier suppression intensity for the singly charged ions. Often, much higher driving intensities, $\sim10^{15}\:$W/cm$^2$, are reported \cite{Ganeev2005boron}. Only in these conditions, strong \textsc{xuv} emission had been observed and at lower intensities, the generated photon number drops very rapidly \cite{Elouga2007Influence}. 

The limitation in our experiments to an intensity $\lesssim2\times10^{14}\:$W/cm$^2$ prevented us from exploring the full parameter range typically used to optimize \textsc{hhg} in ablation plasma plumes, which is probably one of the reasons that lead to a very low level of \textsc{xuv} signal in our experiments with all targets we have used. Deteriorated phase matching in the \textsc{hhg} medium due to the strong wavefront distortions of the \textsc{hhg} driving beam will be another reason. A further limitation was imposed by the minimum distance of the generating beam to the target surface. \stef{When the target surface was placed too close to the generating beam axis (e.g. at the optimum of 100~\textmu m found in ref. \cite{Ganeev2006Harmonic}), we found part of the annular generating beam to be deflected on axis so that its energy was not completely blocked by the iris before the focusing into the MBES (cp. figure \ref{Fig:setup}). This effect could be minimized by increasing the distance of the target surface from the generating beam axis to $\approx300\:$\textmu m, at which point the refracted \textsc{ir} light from the generating beam did no longer generate sidebands on its own. Its intensity was thus considerably weaker than that of the \textsc{ir} probe beam in the MBES interaction volume. The origin of the refraction is probably a steep electron density gradient close to the target surface, which levels off with increasing distance. Increasing the generating beam--target distance from 100 to 300~\textmu m resulted in a decrease by a factor of $\approx2$ in the generated \textsc{xuv} emission, which is most likely due to a rapid drop of the ion density with the distance from the target surface. This drop could possibly have been compensated somewhat by increasing the delay between heating pulse and generating pulse, so as to give more time to the ablated ions to propagate over the increased distance to the generating beam axis.}
In total, we thus were far from the $10^{-5}$ conversion efficiency reported in \cite{Ganeev2005boron,Ganeev2005Silver} for ablation plasma harmonics. In our case, the signal was one to two orders of magnitude weaker than that obtained with argon gas as \textsc{hhg} medium under otherwise equal conditions. This leads us to estimate the conversion efficiency in our experiments as $\sim10^{-8}$.

The low level of signal could hardly be dealt with by averaging more laser shots per photoelectron spectrum since the metal targets were not moved during data acquisition, i.e. the picosecond \stef{heating} pulse hit the surface always at the same spot. Very little material is actually ablated per laser shot and most targets could easily stand $\sim1000$ shots without any degradation of the generated \textsc{xuv} signal. For a cross-correlation measurement, however, we typically use $8000$ laser shots (80 delay steps and averaging 100 shots per spectrum), so there was really no margin left to increase the averaging. On the contrary, the IR-XUV delay scans had to be made rather quickly and were always accompanied by a gradual decline of signal during the scans.

All data were obtained by ionizing argon gas in the MBES which provides a relatively high photoionization cross-section. Measurements with neon as a detection gas, which has a higher cross-section than argon for photon energies $\geq35\:$eV, proved that no harmonics beyond this value have been emitted by any of the used targets.

\section{Non-resonant Harmonic Emission}
\label{sec:plasmanonreso}

\stef{We have studied three different targets for non-resonant \textsc{HHG} in plasmas:} lead ($I_\mathrm{p}(\mathrm{Pb})=7.4\:$eV, $I_\mathrm{p}(\mathrm{Pb^+})=15.0\:$eV, $I_\mathrm{p}(\mathrm{Pb^{2+}})=31.9\:$eV), gold ($I_\mathrm{p}(\mathrm{Au})=9.2\:$eV, $I_\mathrm{p}(\mathrm{Au^+})=20.5\:$eV) and chrome ($I_\mathrm{p}(\mathrm{Cr})=6.8\:$eV, $I_\mathrm{p}(\mathrm{Cr^+})=16.5\:$eV, $I_\mathrm{p}(\mathrm{Cr^{2+}})=31.0\:$eV). The neutral atoms will in all three cases not contribute significantly to the high harmonic emission due to their low ionization potentials, $I_\mathrm{p}$, leading to extremely low barrier suppression intensities $I_\mathrm{BS}\approx 1\times10^{13}\:$W/cm$^2$. The doubly charged ions, on the other hand have $I_\mathrm{BS}\approx4\times10^{14}\:$W/cm$^2$, and we probably do not come sufficiently close to this value with the experimental intensity. It should thus be the singly charged ions that give the dominating contribution to \textsc{hhg} in all three cases: $I_\mathrm{BS}[\mathrm{Pb}^+]=5\times10^{13}\:$W/cm$^2$, $I_\mathrm{BS}[\mathrm{Au}^+]=1.7\times10^{14}\:$W/cm$^2$, and $I_\mathrm{BS}[\mathrm{Cr}^+]=7\times10^{13}\:$W/cm$^2$. At least for lead and chrome we should reach effective intensities in the plasma plume of several $I_\mathrm{BS}$.

Ganeev et al. \cite{Ganeev2005Silver}, who have studied \textsc{hhg} with a silver target ($I_\mathrm{p}(\mathrm{Ag})=7.6\:$eV, $I_\mathrm{p}(\mathrm{Ag^+})=21.5\:$eV, $I_\mathrm{p}(\mathrm{Ag^{2+}})=34.8\:$eV)) also conclude on dominant emission from singly charged ions. They observe a saturation of the cut-off position at harmonic 57 when increasing the \textsc{hhg} driving pulse energy above a value corresponding to an estimated intensity of $\approx4\times10^{14}\:$W/cm$^2$ (a factor 2 above $I_\mathrm{BS}[\mathrm{Ag}^+]$). This harmonic order corresponds to the cut-off position expected for an emission dominated by the singly charged ions and the saturation could be explained by two effects: the free electrons created by ionizing Ag$^+$ leading to self-defocusing of the generating beam and to depletion of the ion ground state. We have also used a silver target in our experiments. Although its second ionization potential  is very similar to that of gold, we have observed, in contrast to gold, only an extremely weak harmonic emission that did not allow performing any form of temporal characterization. This points at additional reasons for the low harmonic yield beyond a simply too low driving laser intensity. It turns out that high conversion efficiencies are not achieved so easily after all and that very intense laser pulses with excellent wave front quality are necessary.

\begin{figure}[tb]
	\centering
	\includegraphics[width=0.28\textwidth]{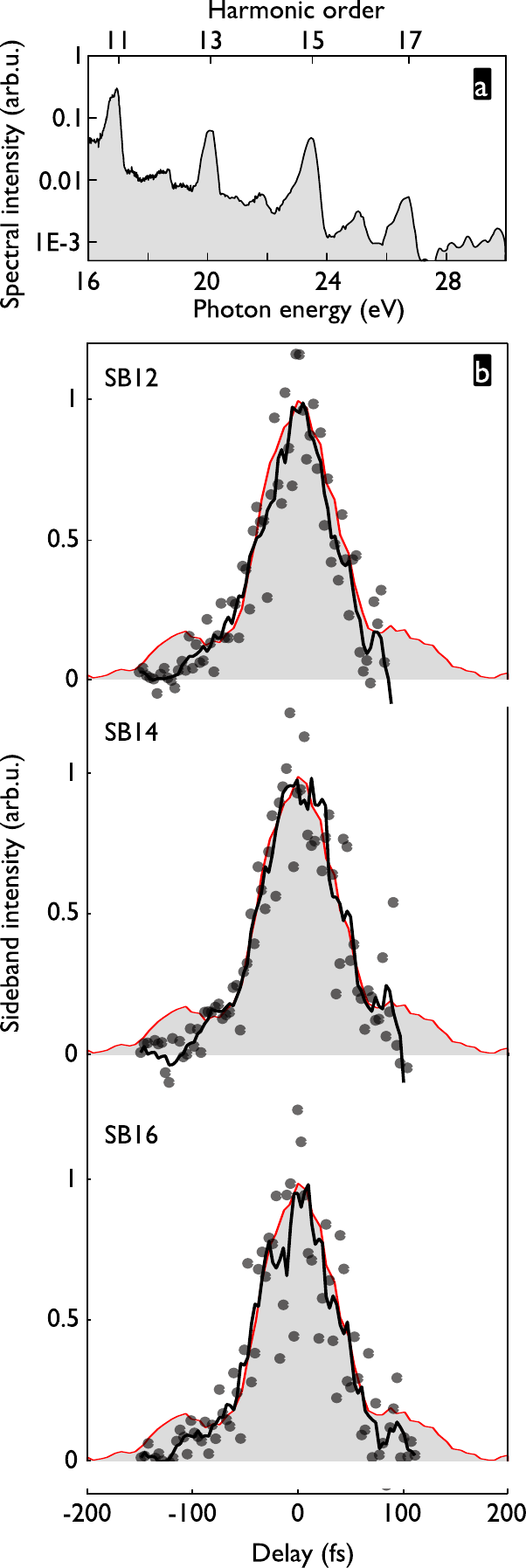}
	\caption{Lead target. (a) Spectral intensity, obtained by summing up photoelecton spectra of the cross-correlation scan and correcting for the argon photoionization cross-section. The quantity shown is thus proportional to the generated \textsc{xuv} spectral intensity, except for the fact that sidebands are visible, which of course only occur in the photoelectron wave packets. (b) Intensities of sidebands 12, 14 and 16 as a function of the XUV-IR delay. Dots mark the raw data points and the black line is a 5-point running average. The grey shaded area and red line show a typical auto-correlation trace of the \textsc{ir} laser.}
	\label{fig:Pb}
\end{figure}

\begin{figure}[tb]
	\centering
	\includegraphics[width=0.28\textwidth]{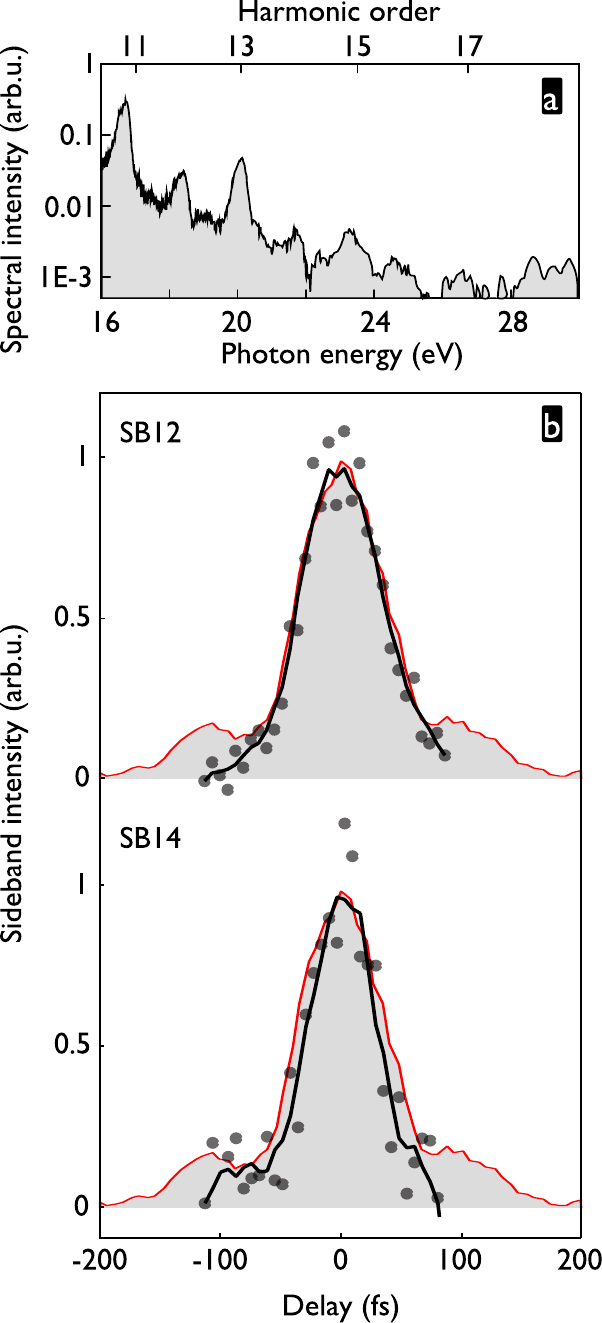}
	\caption{Gold target. (a) Spectral intensity, obtained by summing up photoelecton spectra of the cross-correlation scan and correcting for the argon photoionization cross-section. The quantity shown is thus proportional to the generated \textsc{xuv} spectral intensity, except for the fact that sidebands are visible, which of course only occur in the photoelectron wave packets. (b) Intensities of sidebands 12 and 14 as a function of the XUV-IR delay. Dots mark the raw data points and the black line is a 3-point running average. The grey shaded area and red line show a typical auto-correlation trace of the \textsc{ir} laser.}
	\label{fig:Au}
\end{figure}

\begin{figure}[tb]
	\centering
	\includegraphics[width=0.28\textwidth]{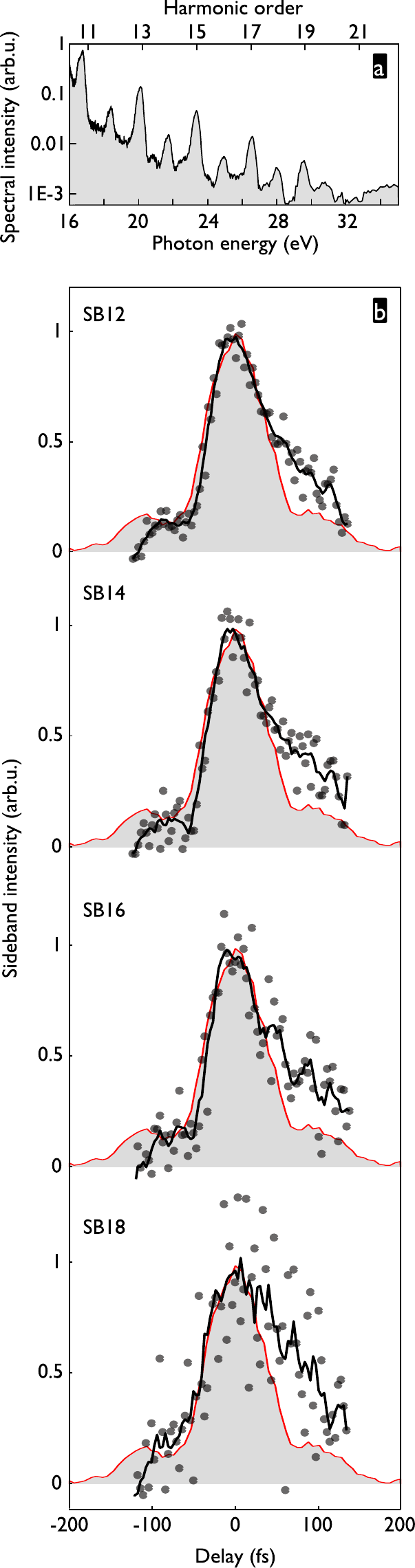}
	\caption{Chrome target. (a) Spectral intensity, obtained by summing up photoelecton spectra of the cross-correlation scan and correcting for the argon photoionization cross-section. The quantity shown is thus proportional to the generated \textsc{xuv} spectral intensity, except for the fact that sidebands are visible, which of course only occur in the photoelectron wave packets. (b) Intensities of sidebands 12 to 18 as a function of the XUV-IR delay. Dots mark the raw data points and the black line is a 5-point running average. The grey shaded area and red line show a typical auto-correlation trace of the \textsc{ir} laser.}
	\label{fig:Cr}
\end{figure}

Harmonic spectra generated with lead, gold, and chrome targets are shown in figures \ref{fig:Pb}a, \ref{fig:Au}a and \ref{fig:Cr}a, respectively. \stef{The value of $I_\mathrm{p}(\mathrm{Pb^+})$ and the corresponding $I_\mathrm{BS}[\mathrm{Pb}^+]$ suggest a maximally achievable cut-off position at harmonic 15 for lead, in agreement with our observation. The same is true for the chrome target, where $I_\mathrm{p}(\mathrm{Cr^+})$ and $I_\mathrm{BS}[\mathrm{Cr^+}]$ lead to an expected cut-off at harmonic 19, as observed. We thus find, in contrast with earlier studies \cite{Ganeev2007}, that the \textsc{hhg} cut-off does not further increase once the barrier suppression intensity is reached.} For gold, $I_\mathrm{BS}[\mathrm{Au}^+]$ should allow at least a cut-off at harmonic 33. It may be due to a particularly low effective intensity in the measurement shown in figure \ref{fig:Au} that  no harmonics above the 15th order could be observed with gold.

Measured cross-correlation traces for lead, gold and chrome samples are shown in figures \ref{fig:Pb}b, \ref{fig:Au}b and \ref{fig:Cr}b, respectively. Positive delays correspond to the \textsc{ir} probe pulse preceding the \textsc{xuv} pulse. All traces have been normalized \emph{(i)} by the spectrally integrated signal for all XUV-IR delays to remove signal fluctuations and \emph{(ii)} by the average value of the three highest points around zero delay, thus restricting all cross-correlation traces to $[0,1]$.

For all cases, the results are compared to a typical auto-correlation trace of the \textsc{ir} probe pulse. Such auto-correlation traces are routinely taken at the exit of the laser compressor and hardly vary from day to day. We have also done SPIDER measurements \cite{Walmsley2009tutorial} of the \textsc{ir} probe pulses after the MBES, i.e. after propagation through the whole optical setup. These measurements revealed that the \textsc{ir} pulse shape was not a simple Gaussian: a small pre-pulse preceded the main laser pulse by a few 10 fs, in
agreement with the pedestal visible in the auto-correlation trace. Higher-order phase terms acquired during propagation through the long delay line could be responsible for this. Otherwise, no significant pulse distortions were found---whether there was an ablation plasma plume to be traversed or not. It should thus be justified to always compare to the same ``standard'' auto-correlation curve.

For lead, the measured cross-correlation traces very closely resemble the \textsc{ir} laser auto-correlation, both in shape and width $\tau_\mathrm{SB}\approx(80\pm10)\:$fs FWHM. The same is true for sideband 12 in the case gold gold, while sideband 14 has a reduced width of $\tau_\mathrm{SB}\approx(70\pm10)\:$fs FWHM. The measurements with the chrome target were made in a different data acquisition run and present a strong asymmetry and a significantly increased width of the XUV-IR cross-correlation as compared to the ``standard'' \textsc{ir} auto-correlation. A clear distortion of the traces appears at positive delays, where the \textsc{xuv} overlaps with the falling edge of the \textsc{ir}. Therefore, it is most likely that the reason
for the distortion is a longer and asymmetric \textsc{ir} probe pulse, possibly with a post-pulse, in this particular run rather than any effect specific to the chrome target.

\section{Resonance-Enhanced Harmonic Emission}
\label{sec:plasmareso}
The probably most interesting aspect of ablation plasmas as an \textsc{hhg} medium is radiative transitions with large oscillator strengths in ions which have been shown to lead to strong enhancement of isolated harmonic orders with photon energy resonant with the transition energy.  We have studied two targets where such resonances exist in the singly charged ions: indium ($I_\mathrm{p}(\mathrm{In})=5.8\:$eV, $I_\mathrm{p}(\mathrm{In^+})=18.9\:$eV, $I_\mathrm{p}(\mathrm{In^{2+}})=28.0\:$eV) and tin ($I_\mathrm{p}(\mathrm{Sn})=7.3\:$eV, $I_\mathrm{p}(\mathrm{Sn^+})=14.6\:$eV, $I_\mathrm{p}(\mathrm{Sn^{2+}})=30.5\:$eV). These are also the two targets used in \cite{Ganeev2006Strong,Suzuki2006Anomalous}. The transitions to which the observed effect is attributed, are \mbox{4d$^{10}$ 5s$^2$ $^1$S$_0$ $\rightarrow$  4d$^9$ 5s$^2$ 5p ($^2$D) $^1$P$_1$} in In$^+$ at 19.92 eV, and \mbox{4d$^{10}$ 5s$^2$ 5p $^2$P$_{3/2}$ $\rightarrow$  4d$^9$ 5s$^2$ 5p$^2$ ($^1$D) $^2$D$_{5/2}$} in Sn$^+$ at 26.27 eV. These two transitions have also been observed in photoabsorption spectroscopy \cite{Duffy2001In,Duffy2001}, and expectionally large $gf$ values\footnote{The $gf$ value is the product of the oscillator strength $f$ of a transition and the statistical weight $g$ of the lower level.} have been calculated. 

The transition energies for the indium and tin resonances are not exactly resonant with the 13th and 17th harmonic of a 793 nm laser ($\hbar\omega_0=1.564\:$eV), respectively, but they are likely driven into resonance by an AC Stark shift in the strong laser field \cite{Ganeev2006Strong,Suzuki2006Anomalous}. 
The mechanism of resonance-enhancement has not yet been clearly identified and it is currently a hot topic of both fundamental and applied interest \cite{Milosevic2007resonant,Elouga2008Correlation,Kulagin2009,Milosevic2010,Strelkov2010,Redkin2010,Frolov2010Potentialbarrier,Tudorovskaya2011resonance}. 
Advanced characterization of the emission would give valuable input to theories describing the role of resonances in strong-field processes. \stef{The model that is most successful in reproducing experiments is a four-step model by Strelkov \cite{Strelkov2010}, also supported by numerical TDSE solutions by Tudorovskaya and Lein \cite{Tudorovskaya2011resonance}. It splits the last step of the usual \textsc{hhg} process into two: the recolliding electron is first captured into the resonance and then relaxes to the ground state emitting the \textsc{xuv} photon. The resonance is thus populated by recolliding electrons and then radiates according to its lifetime. For long lifetimes, this should lead to resonant \textsc{xuv} emission even after the driving laser pulse and other finer temporal features due to interference of newly captured electrons with already existing population of the resonance from the preceding half-cycles \cite{Tudorovskaya2011resonance}. Thus, temporal and/or spectral phase measurements are of particular relevance to the fundamental question of how the resonant enhancement really proceeds.  Likewise, the assessment of the potential for very intense \textsc{xuv} pulses with ultrashort, possibly attosecond, duration requires temporal characterization measurements.}

\begin{figure}[tb]
	\centering
	\includegraphics[width=0.28\textwidth]{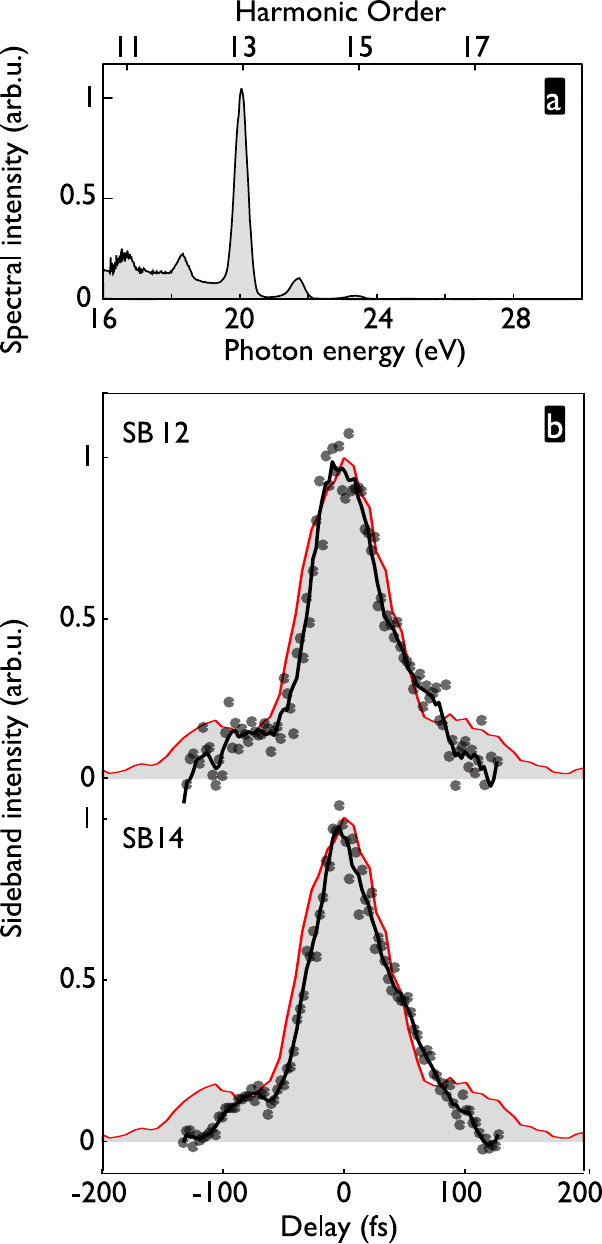}
	\caption{Indium target. (a) Spectral intensity, obtained by summing up photoelecton spectra of the cross-correlation scan and correcting for the argon photoionization cross-section. The quantity shown is thus proportional to the generated \textsc{xuv} spectral intensity, except for the fact that sidebands are visible, which of course only occur in the photoelectron wave packets. (b) Intensities of sidebands 12 and 14 as a function of the XUV-IR delay. Dots mark the raw data points and the black line is a 5-point running average. The grey shaded area and red line show a typical auto-correlation trace of the \textsc{ir} laser.}
	\label{fig:indium}
\end{figure}
\begin{figure}[tb]
	\centering
	\includegraphics[width=0.28\textwidth]{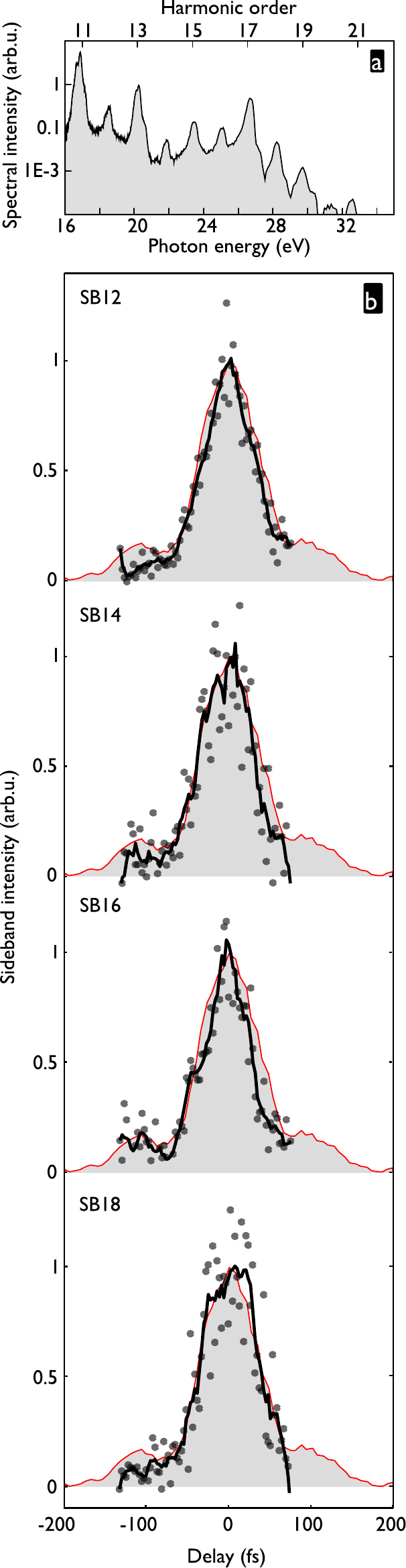}
	\caption{Tin target. (a) Spectral intensity, obtained by summing up photoelecton spectra of the cross-correlation scan and correcting for the argon photoionization cross-section. The quantity shown is thus proportional to the generated \textsc{xuv} spectral intensity, except for the fact that sidebands are visible, which of course only occur in the photoelectron wave packets. (b) Intensities of sidebands 12 to 18 as a function of the XUV-IR delay. Dots mark the raw data points and the black line is a 5-point running average. The grey shaded area and red line show a typical auto-correlation trace of the \textsc{ir} laser.}
	\label{fig:tin}
\end{figure}

Harmonic spectra generated with indium and tin targets are shown in figures \ref{fig:indium}a and  \ref{fig:tin}a, respectively. For indium, harmonic 13 completely dominates the spectrum and is actually generated with an efficiency comparable to that obtained with rare gas atoms in our setup. This is still orders of magnitude away from the $\sim10^{-4}$ efficiency reported in \cite{Ganeev2006Strong}, but it is a clear sign of a very efficient enhancement process. For tin, the intensity of the resonant harmonic 17 is clearly dominant over the neighboring orders, but still a factor 2 smaller than that of harmonic 13. The resonant enhancement process in our experiments with tin is thus less efficient than in that of \cite{Suzuki2006Anomalous}. The driving laser wavelength in this run was slightly shorter in our case (793~nm as compared to 795~nm) and harmonic 17 was less resonant in our conditions.

However, in a different experimental run, we increased the laser wavelength to 796~nm, which resulted in more resonant conditions \cite{Haessler2011submitted} and equal intensities of harmonics 13 and 17. This is still far from the overwhelming dominance of harmonic 17 in \cite{Suzuki2006Anomalous}, which may be due to different plasma conditions in our experiment, caused by, e.g., the mentioned relatively large  ($\approx300\:\mu$m) target distance and a smaller \stef{heating} pulse spot size. Poor phase matching due to the mentioned driving \textsc{ir} wavefront distortions as well as a lower effective driving intensity may be additional reasons. Finally, the pulse duration of 150~fs was much longer in \cite{Suzuki2006Anomalous} than here.

Results of cross-correlation scans analogous to those described in section \ref{sec:plasmanonreso} are shown in figures \ref{fig:indium}b and \ref{fig:tin}b for the indium and tin target, respectively and a 793-nm laser wavelength. For indium, it is reasonable to assume that the sideband intensity involves essentially one contribution, namely that of a quantum path ``harmonic 13 $\pm$ one \textsc{ir} photon''. Both sidebands should thus give cross-correlation traces looking perfectly alike, which is confirmed in the experiment. Thanks to the fairly strong harmonic~13 emission, the signal-to-noise ratio was much better than for other cross-correlation measurements presented here. We find a width of $\tau_\mathrm{SB}\approx(73\pm5)\:$fs FWHM.

The spectrum obtained with the tin target has the advantage that we can compare sidebands involving contributions from non-resonant harmonics (sidebands 12 and 14) to sidebands involving the resonance-enhanced harmonic (sidebands 16 and 18). Both behave essentially in the same way: $\tau_\mathrm{SB}\approx(75\pm10)\:$fs FWHM. Note that the cross-correlations measured for the tin target at a 796-nm laser wavelength (shown in \cite{Haessler2011submitted}) were very similar to the ones obtained at 793~nm despite the more resonant conditions.

In general, the cross-correlation traces obtained here resemble very much those obtained for non-resonant harmonics, namely having only a slightly smaller width than the laser auto-correlation trace and, with a small shoulder at negative delays, showing a signature of a pre-pulse in the \textsc{ir} probe pulse profile. \stef{We thus do not find a significant increase of the pulse duration of the resonance-enhanced harmonics.}

\section{Discussion}
\label{sec:discussion}

Although the acquired traces have a reasonably good signal-to-noise ratio and have been been very well reproducible in several scans on the same day, the quantitative result that can be inferred from them will bear a rather large uncertainty. For one thing, equation \ref{eq:plasma:Xcotau} can only give an estimate for the \textsc{xuv} pulse duration since it is based on Gaussian envelopes, and we know that the \textsc{ir} probe pulse was not precisely Gaussian-shaped. A full de-correlation is in principle possible if the precise \textsc{ir} pulse shape is known, but this requires cross-correlation data of exceptionally high quality and is not applicable here. For another thing, using equation \ref{eq:plasma:Xcotau} and considering $\tau_\mathrm{IR}$ and $\alpha$ to be known exactly, the error, $\Delta\tau_\mathrm{XUV}$, passed on to $\tau_\mathrm{XUV}$ from an experimental uncertainty $\Delta\tau_\mathrm{SB}$ is by a factor $\tau_\mathrm{SB}/\sqrt{\tau_\mathrm{SB}^2-\tau_\mathrm{IR}^2/\alpha}$ larger than that measurement error. When the \textsc{ir} probe pulse is of similar duration or longer as the \textsc{xuv} pulse to be probed, this factor becomes rather large, which leads to the large errors in the values for the harmonic pulse durations given below.

We exclude systematic experimental errors from contributing to the uncertainties given for the widths of the cross-correlation traces. The relation between piezo-steps and time delay was verified by measuring the frequency $\omega_\mathrm{IR}$ of the total signal modulation in scans with interferometric stability. Moreover, a delay jitter larger than our step size of typically 3.3 fs induced by an exaggerated blurring of the interferences by means of the air conditioning seems very unlikely. Our experimental error thus essentially owes to limited signal-to-noise and relatively low statistics.


In order to apply  equation \ref{eq:plasma:Xcotau}, we need to determine the value of $\alpha$ for our experimental conditions. From the measured relative intensities of harmonic and sideband peaks at maximum overlap, as well as from the very small ($(50\pm10)\:$meV) ponderomotive shift measured for the harmonic \stef{11} peak in all cross-correlation measurements \stef{(at the lowest photo-electron energies, the spectral resolution is best)}, we estimate a peak intensity of: $I_\mathrm{IR}^\mathrm{probe}=(8\pm2) \times10^{11}\:$W$\:$cm$^{-2}$. For our photo-electron energy range and the intensity range of (4--8)$\times10^{11}\:$W cm$^{-2}$, relevant to our experiment, the exact expression for $f[I_\mathrm{IR}]$, given as equation (4) in \cite{Bouhal1997}, is well approximated by a power law with exponent $\alpha=0.8$.

		
\stef{With equation \ref{eq:plasma:Xcotau},} we thus find for all observed harmonics from the lead target: $\tau_\mathrm{XUV}=(50\pm15)\:$fs. The same value can be given for the average duration of harmonics 11 and 13 from the gold target, whereas the average duration of harmonics 13 and 15 is found as $\tau_\mathrm{XUV}=(35\pm15)\:$fs. As for the chrome target, the shape of the cross-correlations, shown in figure \ref{fig:Cr}, clearly reveals a significant deviation from Gaussian envelopes. Above the half-maximum, they do, however, present a width and shape very similar to the ``standard \textsc{ir} auto-correlation curve'' shown as a comparison, so that we tend to make the same conclusion as for the lead target: $\tau_\mathrm{XUV}=(50\pm15)\:$fs.

For the resonance-enhanced harmonic 13 from indium, we find: $\tau_\mathrm{XUV}=(40\pm10)\:$fs. Similarly, we find for all harmonic orders from the tin target, resonance-enhanced or not, a duration of $\tau_\mathrm{XUV}=(45\pm10)\:$fs.

For a 55-fs driving pulse, these pulse durations are rather long. Plateau harmonics generated in neutral gases with an intensity well below barrier suppression have been found to have about half the duration of the driving pulse \cite{Mairesse2005HHSPIDER}. Close to the cutoff, harmonics should be even shorter. A very likely reason is that in our experiments, \textsc{hhg} was saturated before the driving pulse peak. Indeed, in a saturated medium where rapid ionization due to strong barrier suppression depletes the medium ground state, the harmonic yield has a very slow or even flat dependence on the driving laser intensity, such that the \textsc{xuv} pulse envelope may even have a larger FWHM duration than that of the \textsc{ir} driver, as reported in, e.g., \cite{Glover1996}. This may indeed explain the relatively long durations measured for the lead, chrome and tin targets, where the observed \textsc{hhg} cutoff at H17--H19 and the classical cutoff law \cite{Schafer1993Above,Corkum1993Plasma} suggest an effective driving intensity of only $0.5\times10^{14}\:$W$\:$cm$^{-2}$, which also corresponds to the barrier suppression intensity for the singly-charged ions. For gold and indium, which have higher ionization potentials and thus barrier suppression intensities, we have indeed measured slightly shorter durations of the high harmonic emission, consistent with a less pronounced saturation effect.

\stef{For a conclusion with regard to the mentioned four-step model for resonance-enhanced \textsc{hhg}, we would need to know the lifetime of the relevant resonances in In$^+$ and Sn$^+$. The resonance lifetime in a strong laser field is not well known, but taking the field-free lifetime as an indication ($\sim20\:$fs for both indium and tin, when estimating resonance widths of $<0.04\:$eV from the data shown in \cite{Duffy2001In,Duffy2001}), we would expect a significantly increased emission duration. Our experimental findings are thus unexpected, but could be reconciled with the four-step model if the resonance lifetime in a strong laser field (and in the presence of resonant $\textsc{xuv}$ photons) is significantly decreased. This did not seem to be the case in the numerical 1D TDSE solutions of \cite{Tudorovskaya2011resonance}, where a shape resonance has been considered because the calculation was limited to a single active electron. The ``real'' resonances are auto-ionizing states and thus of multi-electron nature, and of course occur in 3D space. Their reaction to external fields may thus be very different.} 

\stef{While our results are not yet conclusive with regard to the above questions, they give new experimental input to the theoretical effort of modeling resonance-enhanced HHG. Regarding the applied aspect, our measurements show that resonance-enhanced harmonics are an excellent potential source of highly intense \textsc{xuv} pulses with femtosecond duration.}

\section{Conclusions}

These results constitute the first temporal characterization of the femtosecond envelope of the high-order harmonic emission from ablation plasma plumes. The complex nature of this medium containing different kinds of ions and a rather high free electron density does not allow relying on straightforward analogies with the well known \textsc{hhg} in neutral gases. \stef{The confirmation, found in our results, that the \textsc{xuv} emission from plasmas, both resonant and non-resonant, has a femtosecond envelope thus constitutes an important advance.} While the determined harmonic pulse durations bear significant relative uncertainties, \stef{we consistently find \textsc{xuv} pulse durations that are shorter than the driving  laser pulse for all plasma targets.} 

\stef{While our results for the resonance-enhanced harmonics are not yet conclusive with regard to Strelkov's four-step model, they give new experimental input to the theoretical effort of modeling.}

The surprising similarity of the pulse envelope for resonant and non-resonant harmonics is a very good news which, in view of the very high conversion efficiencies of $10^{-4}$ observed in other experiments, opens the perspective of a very high peak flux tabletop \textsc{xuv} source. The central wavelength of these pulses could be selected by choosing a target material with a strong transition at the desired energy. 
	
Our measurements could be complemented by more advanced temporal characterization. For instance, in so-called XFROG measurements, using a shorter \textsc{ir} probe pulse, such that $\tau_\mathrm{XUV}>\tau_\mathrm{IR probe}$, would allow a more precise determination of the \textsc{xuv} pulse duration as well as a measurement of the harmonic chirp rate, as demonstrated in \cite{Norin2002,Mauritsson2004xfrog,Mauritsson2005}. Alternatively, a high-harmonic SPIDER measurement would give direct access to both temporal envelope and phase \cite{Mairesse2005HHSPIDER}. Finally, a \emph{complete} characterisation of the harmonic emission including its attosecond (sub)-structure with the FROG-CRAB method \cite{Mairesse2005} would of course be the most desirable progress, albeit certainly also the most challenging one. All these would give information on the intensity dependence of the harmonic phase, which may be different for resonant and non-resonant harmonics, thus shedding more light on the mechanism behind resonant enhancement.

\ack
We acknowledge financial support from the EU-FP7-ATTOFEL, ANR-09-BLAN-0031-01. S.H. acknowledges support through the fellowship M1260-N16 of the Austrian science fund (FWF) as well as the Marie-Curie fellowship EU-FP7-IEF-MUSCULAR.

\bibliographystyle{iopart-num}
\bibliography{stefan}

\end{document}